# An optimization framework for route design and allocation of aircraft to multiple departure routes

V. Ho-Huu[*], S. Hartjes, H. G. Visser, R. Curran

*Faculty of Aerospace Engineering, Delft University of Technology, P.O. Box 5058, 2600 GB, Delft, The Netherlands*

E-mails: v.hohuu@tudelft.nl  (V. Ho-Huu)
s.hartjes@tudelft.nl  (S. Hartjes)
h.g.visser@tudelft.nl (H. G. Visser)
r.curran@tudelft.nl   (R. Curran)
[*]Corresponding author: V. Ho-Huu, v.hohuu@tudelft.nl; vinh.ho.h@gmail.com

**Abstract**

In this article, we present the development of a two-step optimization framework to deal with the design and selection of aircraft departure routes and the allocation of flights among these routes. The aim of the framework is to minimize cumulative noise annoyance and fuel burn. In the first step of the framework, multi-objective trajectory optimization is used to compute and store a set of routes that will serve as inputs in the second step. In the second step, the selection of routes from the set of pre-computed optimal routes and the optimal allocation of flights among these routes are conducted simultaneously. To validate the proposed framework, we also conduct an analysis involving an integrated (one-step) approach, in which both trajectory optimization and route allocation are formulated as a single optimization problem. A comparison of both approaches is then performed, and their advantages and disadvantages are identified. The performance and capabilities of the present framework are demonstrated using a case study at Amsterdam Airport Schiphol in The Netherlands. The numerical results show that the proposed framework can generate solutions which can achieve a reduction in the number of people annoyed of up to 31% and a reduction in fuel consumption of 7.3% relative to the reference case solution.

**Keywords**: departure routes; trajectory optimization; aircraft allocation; noise abatement; aircraft noise; airport noise.

# 1. Introduction

Over the past decades, aircraft noise and pollutant emissions have remained major issues in the aviation sector. These environmental issues do not only negatively affect the quality of life of communities surrounding airports and global climate change, but also hamper the expansion of flight and airport operations. Research efforts towards mitigating these negative impacts have received much attention, and some important achievements have been reported (Casalino et al., 2008; Lee et al., 2009). However, it appears that these attempts still remain insufficient to meet the need to accommodate the expected rapid growth in air traffic demand in the coming years (Girvin, 2009). In order to support the sustainable development of the aviation industry, more research on these topics is necessary. There are several possible strategies to achieve the objective, for example, adjusting operational procedures at airports, developing new aircraft technologies and using alternative fuels, or setting new rules and regulations (Marais et al., 2013). While new technologies and sustainable fuels may provide a significant reduction in environmental impact, they require more effort and time to develop and implement. In contrast, despite having smaller mitigation potential, operational changes can be carried out in the short-term period (Marais et al., 2013). As promising options in this category, the design of optimal aircraft routes and the assignment of aircraft to specific runways and routes have been well recognized and have shown promising results over the years (Frair, 1984; Visser, 2005).

Regarding the design of optimal aircraft routes, research has been typically aimed at minimizing community noise impact and fuel consumption or pollutant emissions such as nitrogen oxides ($NO_x$) and carbon dioxides ($CO_2$). Based on the use of noise criteria, studies can be classified into two groups. The first group includes research which employs noise criteria derived from a single fly-over noise event. For instance, the awakening criterion proposed by the Federal Interagency Committee on Aviation Noise (FICAN, 1997) was used in Refs. (Hartjes et al., 2010; Hartjes and Visser, 2016; Hogenhuis et al., 2011; Visser and Wijnen, 2003, 2001), while the more recent dose-response relationship developed by the American National Standards Institute (ANSI, 2008) was utilized in Refs. (Ho-Huu et al., 2017; Yu et al., 2016; Zhang et al., 2018). Also, noise nuisance criteria based on the maximum perceived sound level were applied by Prats et al. (2011, 2010a, 2010b) and Torres et al.

(2011). The second group consists of studies in which the aggregation of multiple noise events is utilized as a noise criterion. For example, the annoyance criterion based on the $L_{den}$ cumulative noise metric was used in Refs. (Braakenburg et al., 2011; Ho-Huu et al., 2018a; Song et al., 2014), whilst a sleep disturbance criterion based on the $L_{night}$ cumulative noise metric was used by Hartjes et al. (2014).

In terms of research on the allocation of flights to specific routes and runways, Frair (1984) developed an integer optimization model to find the optimal allocation of aircraft among available approach and departure routes with the aim of minimizing community annoyance. Kuiper et al. (2012) maximized the number of aircraft movements operating at an airport within an allotted annual noise budget by optimally assigning annual flights to available routes and runways. Kim et al. (2014) minimized airport surface emissions by concurrently allocating aircraft among runways and scheduling departure and arrival flights on these runways. Zachary et al. (2010) formulated and solved an optimization problem to minimize noise and pollutant emissions by simultaneously considering operational procedures, arrival and departure routes, and fleet combination. In later research, with a similar approach, Zachary et al. (2011) evaluated the potential reduction in operational cost that could be gained by optimal solutions. Ganić et al. (2018) and Ho-Huu et al. (2019) developed integer optimization models to allocate flights among available departure and arrival routes with the aim of reducing the population noise exposure, while taking into account daily migrating populations.

A review of the above literature reveals that the design of aircraft routes and the assignment of flights to available routes and runways have indeed been broadly studied. These studies were, however, typically carried out separately, while studies that consider both trajectory optimization and allocation problems at a time are lacking. In particular, with respect to the problem of designing aircraft routes, research has been aimed at finding the optimal flight trajectories for a given standard route. On the other hand, studies focusing on the allocation of aircraft movements generally only considered existing standard routes rather than optimized routes. Consequently, the potential reduction of environmental impact by formulating and solving an integrated optimization problem that consists of these two sub-problems has not yet been fully explored. It is important to note that the two sub-problems of route design and aircraft allocation are intrinsically coupled when a cumulative noise metric such as $L_{den}$ is considered. This coupling is brought about by the fact that the optimal route obtained by the route design problem directly depends on the number of aircraft (of any given types) which is assigned to that route.

Although the problem of integrating both sub-problems could, in principle, be formulated and solved as a single integrated problem, it is likely to be prohibitively large and complicated due to high computational cost. In an attempt to fulfil the above research gaps and to overcome the aforementioned challenges, an optimization framework consisting of two sequential steps is developed in this paper. In the first step, the optimization problem of designing optimal routes is formulated and solved. The results obtained in this step contain sets of optimal routes which can effectively balance between noise annoyance and fuel consumption. Next, these data sets are used as the inputs for the allocation problem in the second step, in which the selection of optimized routes and the optimal allocation of aircraft movements among these routes are conducted concurrently.

In order to assess the reliability of the present approach, we also perform an integrated problem (here also referred to as the one-step approach), which combines both optimization sub-problems into a single integrated problem. A comparison between the one-step and two-step approaches is then presented, and the advantages and disadvantages of both approaches are discussed. Since only the ground tracks of routes were considered in previous research using annoyance criteria (Braakenburg et al., 2011; Hartjes et al., 2014; Ho-Huu et al., 2018a; Song et al., 2014), the potential of including optimized vertical profiles is also evaluated in the proposed framework. As a consequence, two different cases are investigated. In the first case, only the ground tracks are optimized, whilst in the second case, the ground tracks and the vertical profiles are optimized simultaneously. Moreover, the consideration of these two case studies also aims to validate the reliability of the proposed two-step approach, as well as to highlight the drawbacks of the one-step approach. All computational experiments have been carried out in a case study involving Amsterdam Airport Schiphol (denoted as AMS) in The Netherlands.

The remainder of the paper is structured as follows. Theoretical backgrounds are provided in Section 2. Section 3 presents the proposed two-step optimization framework in detail, while numerical results and discussion are presented in Section 4. Finally, conclusions and future work are discussed in Section 5.

## 2. Theoretical background

### 2.1. Aircraft model

To evaluate aircraft performance, an intermediate point-mass dynamic model that has been widely utilized in previous research (Hartjes et al., 2014; Visser and Wijnen, 2001) is employed in this study. This dynamic model relies on the assumptions that: 1) no wind is present, 2) the Earth is flat and non-rotating, 3) flight is coordinated, and 4) the flight path angle is sufficiently small ($\gamma < 15^0$). The equations of motion are then given by

$$\begin{aligned}
\dot{V}_{TAS} &= g_0 \left( \frac{T-D}{W} - \sin\gamma \right), \\
\dot{s} &= V_{TAS} \cos\gamma, \\
\dot{h} &= V_{TAS} \sin\gamma, \\
\dot{W} &= -\dot{m}_0 \, g_0,
\end{aligned} \qquad (1)$$

where $\dot{V}_{TAS}, \dot{s}, \dot{h}, \dot{W}$ are, respectively, the derivatives with respect to time of the true airspeed, ground distance flown, altitude and aircraft weight; and $T$, $D$, $\dot{m}_0$, and $g_0$ are thrust, drag, fuel flow and the gravitational acceleration, respectively.

Since departure operations take place at low airspeeds and altitudes, the equivalent airspeed $V_{EAS}$ can be used as a proxy for the indicated airspeed. Based on the relationship with the true airspeed, $V_{EAS}$ can be defined as follows:

$$V_{EAS} = V_{TAS} \sqrt{\rho/\rho_0}, \qquad (2)$$

where $\rho_0$ and $\rho$ are, respectively, the air density at sea level and the ambient air density.

With the use of the relationship in Eq. (2), Eq. (1) can be redefined by:

$$\begin{aligned}
\dot{V}_{EAS} &= \left[ g_0 \left( \frac{T-D}{W} - \sin\gamma \right) + \frac{1}{2\rho^2} \frac{\partial \rho}{\partial h} V_{EAS}^2 \rho_0 \sin\gamma \right] \sqrt{\rho/\rho_0} \\
\dot{s} &= V_{EAS} \sqrt{\rho_0/\rho} \cos\gamma, \\
\dot{h} &= V_{EAS} \sqrt{\rho_0/\rho} \sin\gamma, \\
\dot{W} &= -\dot{m}_0 \, g_0,
\end{aligned} \qquad (3)$$

in which $\frac{\partial \rho}{\partial h}$ is the derivative of the ambient air density $\rho$ with respect to altitude $h$.

## 2.2. Trajectory parameterization

To parameterize the trajectory of a route, the method presented in Hartjes and Visser (2016) is utilized. This technique divides a trajectory into two different components: a horizontal and a vertical profile. In the horizontal profile, the flight path is constructed by employing Required Navigation Performance (RNP) based on flight legs, relying on two common leg types, i.e., track-to-a-fix (TF) and radius-to-a-fix (RF). This essentially results in a ground track that consists of an alternating sequence of straight segments and constant radius turns. For an example of how these leg types are used to create a route, interested readers can refer to Ho-Huu et al. (2017).

For the generation of the vertical profile, the flight procedures outlined in ICAO (2006) are applied. In the trajectory synthesis conducted in this study, a decrease in altitude and/or a deceleration in velocity during departure is not allowed, and similarly, an increase in altitude and/or acceleration in velocity is also prohibited during approach. The parameterization of the vertical profile has been based on splitting a trajectory into a number of segments. Depending on operational requirements, the two control inputs in each segment, *viz.* the throttle and flight path angle settings, are designated either as design (optimization) variables or their values are directly assigned. The vertical profile is then projected onto the ground track, yielding a complete 3-dimensional trajectory. For more details on the applied technique, interested readers can refer to Refs. (Hartjes and Visser, 2016; Ho-Huu et al., 2017).

By using this technique, the design variables of a route comprise all parameters defining the ground track and vertical profile. However, in case the vertical profile is a priori fixed, the design variables only relate to the parameters defining the ground track.

## 2.3. Optimization criteria

In the field of design and allocation of optimal aircraft routes, two widely used objectives are fuel burn and noise annoyance. While the fuel-burn criterion can be readily assessed by estimating the change of aircraft gross weight during departure, the second criterion is significantly difficult to gauge due to the lack of consistency between single-event and multi-event noise metrics. Two noise criteria that have been broadly utilized in previous studies are the number of expected awakenings (ANSI, 2008) and annoyance (EEA, 2010). The awakening criterion is a single-event noise metric, which only considers a single noise event at a time and hence is only suitable for assessing the noise impact of a single movement of a single aircraft type. Although it might be used to design optimal routes in the first step of the framework, it is not suitable to be used for the allocation problem in the second step, as this step - by definition - considers the impact of multiple aircraft movements. Consequently, it does not represent a feasible option for the proposed framework. The annoyance criterion, however, is based on the accumulation of multiple noise events, and it can, therefore, be applied in both steps. For this reason, this particular criterion has been adopted within the optimization framework. Its implementation is described below.

As indicated by EEA (2010), the percentage of people annoyed (%$PA$) based on the $L_{den}$ cumulative noise metric at a given location on the ground is given by

$$\%PA = 8.588 \times 10^{-6}(L_{den} - 37)^3 + 1.777 \times 10^{-2}(L_{den} - 37)^2 + 1.221\,(L_{den} - 37) \qquad (4)$$

where $L_{den}$ is the day-evening-night noise level, determined as follows:

$$L_{den} = 10\log_{10}\left[\sum_{k \in N_r}\sum_{i \in N_{at}} a_{ki} 10^{\frac{SEL_{ki}+w_{den}}{10}}\right] - 10\log_{10} T \text{ (dBA)}, \qquad (5)$$

where $N_r$ is the total number of departure routes; $N_{at}$ is the total number of aircraft types; $SEL_{ki}$ is the sound exposure level resulted from aircraft type $i$ on route $k$; $w_{den} = \{0,5,10\}$ is a weighting factor to account for day, evening and night time operations; $a_{ki}$ is the number of aircraft type $i$ operating on route $k$; and $T$ is the considered time period in seconds (in this case $T = 24\times3600$ seconds). The $SEL$ metric is calculated at each location on the ground by using a replication of the noise model given in the technical manual of the Integrated Noise Model (INM) (FFA, 2008).

It should be noted that, in order to evaluate this criterion, the total number of movements on each individual route within a given time period needs to be known in advance. However, given that the trajectory optimization process precedes the allocation of flights, the optimal number of flights is yet unknown. To overcome this, a strategy has been developed to identify appropriate estimates of the number of flights that are allocated to each route, allowing to generate a sufficiently comprehensive set of alternative routes in step one. This strategy will be further discussed in Section 4.1.

**2.4. Optimization method**

To solve the optimization problems in two steps, a novel variant of the multi-objective evolutionary algorithm based on decomposition (MOEA/D), recently developed in Ho-Huu et al. (2018a), has been employed. The MOEA/D method was originally proposed by Zhang and Li (2007) and has been proven to be one of the most effective multi-objective evolutionary algorithms in recent years (Trivedi et al., 2016). In MOEA/D, decomposition approaches such as Tchebycheff decomposition are utilized to transform a multi-objective optimization problem into a set of scalar optimization sub-problems. Then, an evolutionary algorithm such as genetic algorithm (GA) or differential evolution (DE) is employed to solve the sub-problems concurrently. The performance of MOEA/D for solving the optimization problems of designing optimal aircraft routes has been clarified in Ho-Huu et al. (2017), demonstrating that MOEA/D performed much better than the non-dominated sorting genetic algorithm II (NSGA-II) proposed by Deb et al. (2002). Since the details of the algorithm have been given in Refs. (Ho-Huu et al., 2018b, 2017; Zhang and Li, 2007) interested readers are encouraged to refer to these references.

**3. A two-step optimization framework**

Before the description of the proposed framework is presented in detail, it is worth mentioning that the first step in the optimization framework (i.e., the design of optimal routes) is a planning step which needs to be executed off-line based on a flight schedule and a variety of runway configurations. Indeed, due to the high computational cost, the first step cannot be executed on-line. Meanwhile, the second step (i.e., the allocation problem) can be performed within half an hour CPU time, and hence it might be quickly adapted to unplanned changes in flight schedules and runway configurations. The details of the two optimization steps are presented below.

**Step 1: design of optimal routes**

The main objective of the first step is to identify optimal routes for a given standard instrument departure route (hereafter referred as SID), in which a trade-off between the number of people annoyed and fuel burn is considered. The optimization problem is formulated as follows:

$$\min_{\mathbf{d}} \quad \left(N_{\text{pa}}(\mathbf{d}), T_{\text{fuel}}(\mathbf{d})\right)$$
$$\text{s.t.} \quad \mu_i(t) \leq \mu_{\max}(h), \quad \forall i \in N_{\text{at}} \tag{6}$$

where $N_{\text{pa}}(\mathbf{d})$ and $T_{\text{fuel}}(\mathbf{d})$ are the two objective functions which are, respectively, the total number of people annoyed and the total fuel burn of all aircraft following the SID, and $\mathbf{d}$ is the vector of design variables that contains the parameters defining the route as described in Section 2. The index $N_{\text{at}}$ is the number of aircraft types, and the variable $\mu_i(t)$ is the bank angle of aircraft type $i$ during a turn. With the use of the assumptions indicated in Section 2.1, the bank angle can be expressed as: $\mu_i(t) = \pm \tan^{-1}\left(\dfrac{V_{\text{TAS},i}^2}{g_0 R}\right)$, where $V_{\text{TAS},i}$ is the true airspeed of aircraft type $i$, and $R$ is the turn radius. The parameter $\mu_{\max}$ is the maximum permissible value of $\mu$, varying according to altitude $h$ (ICAO, 2006).

In Eq. (6), the objective $N_{\text{pa}}(\mathbf{d})$ is calculated by aggregating over all grid cells the product of %$PA$ in Eq. (4) in each grid cell with the population in that cell. The population residing in each grid cell is retrieved from a Geographic Information System (GIS) containing population density data surrounding an airport. It is noted that $N_r$ in Eq. (5) is equal to 1 in this case, since only one SID at a time is evaluated. The objective $T_{\text{fuel}}(\mathbf{d})$ is the sum of the fuel burn of all aircraft during the specific departure period and is evaluated by

$$T_{\text{fuel}}(\mathbf{d}) = \sum_{i \in N_{\text{at}}} a_i \, fuel_i(\mathbf{d}) \tag{7}$$

where $a_i$ is the number of aircraft type $i$, $fuel_i(\mathbf{d})$ is the fuel burn of aircraft type $i$.

By solving the problem defined in Eq. (6) for each SID, the sets of optimal routes and associated performances are found, which then serve as inputs to the optimization problem in the second step. It should be noted that the aircraft types selected when designing optimal routes are assumed to be given.

Also, to be able to adapt to different runway configurations, the optimal routes for all standard routes of each runway should be obtained.

**Step 2: selection of routes and allocation of aircraft to these routes**

Based on the sets of optimal routes obtained in Step 1 for all SIDs, this step aims to define which routes from the sets are preferred, and how many movements of each aircraft type should be allocated to these preferred routes for different operational times. The answer to these questions can be obtained by solving the optimization problem stated as follows:

$$\begin{aligned}
\min_{\mathbf{r},\mathbf{a}} \quad & \left(N_{\text{pa}}(\mathbf{r},\mathbf{a}), T_{\text{fuel}}(\mathbf{r},\mathbf{a})\right) \\
\text{s.t.} \quad & \sum_{k \in SD_s} a_{itk} = T_{\text{at},its}, \ \forall i \in N_{\text{at}}, \ \forall t \in (\text{d,e,n}), \ \forall s \in T_{\text{p}} \\
& \sum_{t \in (\text{d,e,n})} \sum_{i \in N_{\text{at}}} a_{itk} \leq \bar{N}_{\text{f},k}, \ \forall k \in N_r \\
& 0 \leq a_{itk} \leq \bar{a}_{itk}
\end{aligned} \quad (8)$$

where $\mathbf{r} = \{r_1,..,r_k,...,r_{Nr}\}$ is the vector of design variables of departure routes, in which the preferred route $r_k$ is selected from the set of optimal routes $\mathbf{O}_k$ obtained in Step 1 for SID $k$, and $N_r$ is the total number of considered SIDs. The vector $\mathbf{a}$ is the design variable vector of aircraft allocation, in which $a_{itk}$ is the number of aircraft type $i$ at time $t$ on route $k$. The index $t$ is the operating time of aircraft (i.e. day (d), evening (e) or night (n)). The index $s$ is the terminal point (i.e., the end point of departure procedure), and $T_{\text{p}}$ is the set of terminal points. The vector $SD_s$ is the vector that contains SIDs having the same terminal point $s$. The identification of SIDs with the same terminal point allows the algorithm to allocate aircraft movements on different SIDs originating from the same or different runways. The parameter $T_{\text{at},its}$ is the total number of aircraft type $i$ at time $t$ sent to departure routes having the same terminal point $s$. The parameter $\bar{N}_{\text{f},k}$ is the upper bound of the number of movements that route $k$ can handle in a certain period of time. Finally, the parameter $\bar{a}_{itk}$ is the upper bound of the number of aircraft type $i$ on route $k$ at time $t$. It should be noted that the exit point of each flight in a flight schedule can be specified in advance based on its destination airport.

In Eq. (8), the objectives $N_{\text{pa}}(\mathbf{r},\mathbf{a})$ and $T_{\text{fuel}}(\mathbf{r},\mathbf{a})$ are the same as the ones considered in step 1, however, the design variables are different. Specifically, the design variables in this problem represent the selection of routes from the sets of optimal routes for each SID, and the distribution of aircraft on these routes. As in Step 1, the objective $N_{pa}(\mathbf{r},\mathbf{a})$ is evaluated by the sum of the multiplication of %*PA* in each grid cell with the population in that cell. However, the *SEL* metric for each route is now known in advance, as it was stored in Step 1. Consequently, both $L_{den}$ and %*PA* can be determined directly from the *SEL*-data stored in the set of optimal routes by applying Eqs. (5) and (4). The objective $T_{fuel}(\mathbf{r},\mathbf{a})$ is defined as follows:

$$T_{\text{fuel}}(\mathbf{r},\mathbf{a}) = \sum_{k \in N_{\text{r}}} \sum_{i \in N_{\text{at}}} a_{ki}\, fuel_{ik}(r_k) \qquad (9)$$

where $fuel_{ik}(r_k)$ is the fuel burn of aircraft type *i* on route $r_k$.

It should be noted that, in the problem stated in Eq. (8), airspace capacity and aircraft sequence are assumed to be satisfied via the constraint, in which each route can only accommodate a certain number of flights within the considered time frame (24 hours in this case). The actual influence of optimal allocation solutions on the airspace capacity and aircraft sequence, which is a challenging problem, is not considered yet. This aspect will be explored in future work.

## 4. Numerical examples and discussion

In this section, a case study at AMS in The Netherlands, as shown in Fig. 1, is presented to exemplify the capabilities of the proposed framework. For this case study, four existing standard instrument departures (SIDs), *viz.* LEKKO, KUDAD, LUNIX and RENDI departing from two different runways, *viz.* RW18 and RW24, are considered. The SIDs LEKKO and KUDAD both end at the LEKKO intersection, whereas LUNIX and RENDI both terminate at IVLUT. On the selected reference day, 337 flights operating on these routes were recorded. These flights can be classified into three groups with different departure times, *viz.* 237 day flights (07h00-19h00) accounting for 70% of the total traffic volume, 43 evening flights (19h00-23h00) accounting for 13%, and 57 night flights (23h00-7h00) accounting for 17% (Dons, 2012). Although many different aircraft types operate on these routes, for the sake of simplicity all flight movements are assumed to be conducted by either of two aircraft types, namely the Boeing 737-800 (B738) and Boeing 777-300 (B773). It is assumed that the B738 represents all small and medium aircraft, accounting for 80% of the total number of flights, while the B773 represents heavy aircraft accounting for 20%. Both aircraft types are modelled based on the Base of Aircraft Data (BADA) (Nuic et al., 2010). The population data acquired from the Dutch Central Bureau of Statistics (CBS) with a grid cell size of 500×500 m, as shown in Fig. 1, is utilized. The MOEA/D algorithm with a population size of 50 and a maximum number of iterations of 1000 is applied to solve all optimization problems. The simulations are performed on an Intel Core i5, 8GB RAM desktop with the use of MATLAB 2016b.

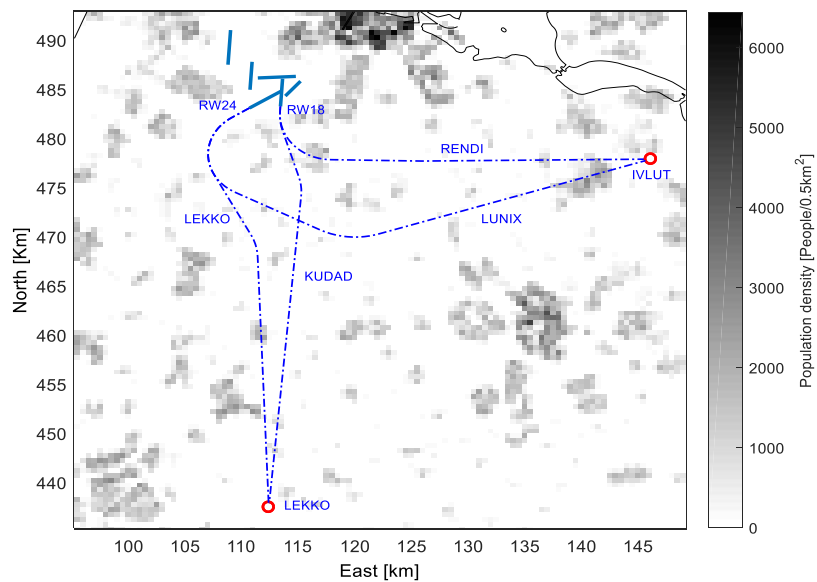

Fig. 1. Illustration of the case study at AMS.

### 4.1. Sensitivity analysis of noise criterion

As mentioned in Section 2.3, to use the noise annoyance criterion for the design of optimal routes, the number of aircraft movements on each SID has to be known a priori. However, this information is unknown within Step 1 of the framework. Therefore, to choose a representative number of movements that leads to a comprehensive set of alternative routes for the allocation problem in the second step, a brief analysis to determine a valid assumption for the number of movements is carried out in this section. For this analysis, the LUNIX SID and the B738 model are used. By considering the actual operational data, approximately 100 flights could be identified, and this number of flights might be used as a "representative" number. However, as the allocation algorithm may actually result in a totally different number of flights assigned to either SID, a significant variation (either positive or negative) to this number needs to be taken into account. Therefore, we assess the results for the assumed number of 50, 100 and 150 flight movements.

The comparison of optimal ground tracks and vertical profiles corresponding to different assumed numbers of flights is illustrated in Fig. 2. Already at first glance, it can be seen from Fig. 2 that the number of flights has a significant influence on the optimal results. The reasons for this are the significant increase of $L_{den}$ with increasing number of movements and the distribution of the population. An illustration of the $L_{den}$ contours for different numbers of flights is given in Fig. 3. By taking a closer look at Fig. 2, however, it can be observed that the solutions obtained by assuming 150 flights include those acquired by assuming 50 and 100 flights. This can also be seen in Fig. 4, where most of the solutions obtained by assuming 50 and 100 flights are on the Pareto front obtained by 150 flights.

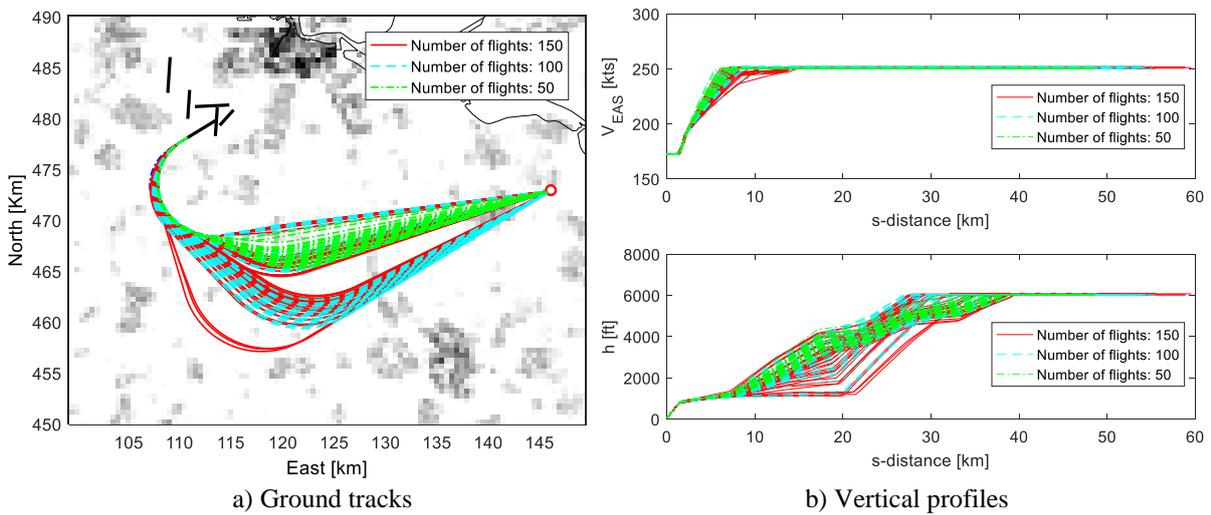

a) Ground tracks        b) Vertical profiles
Fig. 2. Comparison of vertical profiles and ground tracks with different numbers of flights.

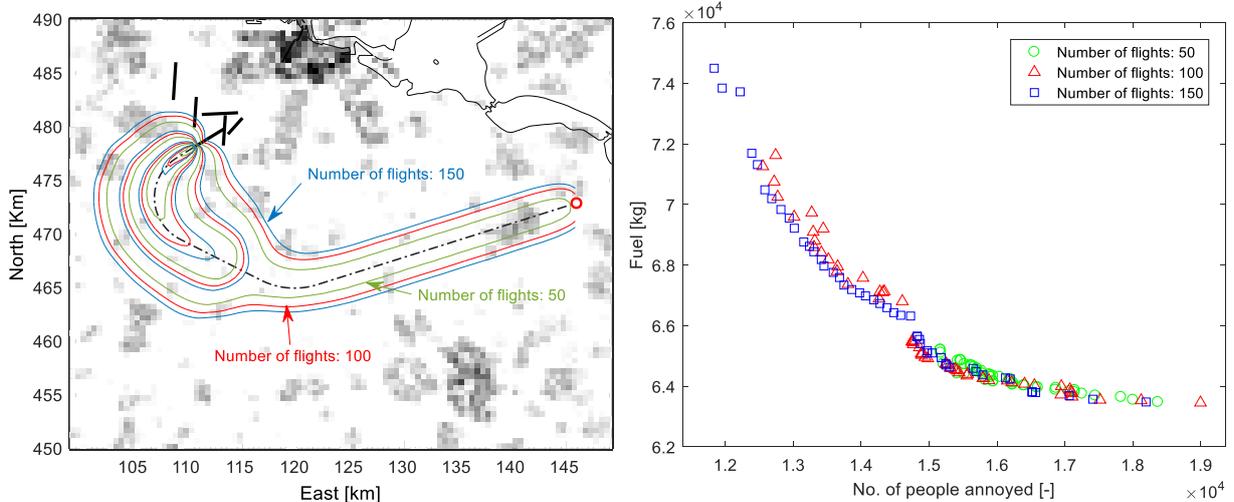

Fig. 3. Comparison of the $L_{den}$ noise contours (37 dBA) caused by different numbers of flights.

Fig. 4. Comparison of objectives with different numbers of flights.

Consequently, assuming a large number of flights results in the most comprehensive set of alternative trajectories. Therefore, the maximum number of flights that can be allocated to a specific SID – based on the projected number of movements in the entire scenario – is considered as an acceptable assumption in Step 1. It should be noted that although only the analysis for the LUNIX SID with a B738 is presented here, the same behavior is also observed for different aircraft types (i.e., B773) and different SIDs.

**4.2. 2D optimization case**

As mentioned before, the problem that integrates both sub-problems can theoretically be formulated and solved. Nevertheless, due to the high computational cost, it is likely to be prohibitively large and complex. To still be able to validate the proposed framework by comparing it with an integrated problem formulation, a relatively simple problem scenario is considered in this section. Specifically, for the optimization problems in the first step, only the ground tracks of routes are optimized, whilst the vertical profiles are fixed and derived from the noise abatement departure procedure 2 (NADP2)[2] (ICAO, 2006) and typical airline procedures.

As shown in Fig. 1, all SIDs are modelled by two turns and three straight legs, which results in 5 design variables for each optimal route design problem. The definition of these design variables can be found in Ho-Huu et al. (2017). It is noted that both B738 and B773 are assumed to follow the same route. The SIDs are assumed to start at the end of the runways at an altitude of 35 ft and a take-off safety speed of $V_2$+10 kts, and to terminate at an altitude of 6000 ft and an equivalent airspeed (EAS) of 250 kts. To design optimal routes for each SID in step one, a number of 150 flights – which is representative for the maximum possible number of movements on any of the SIDs under consideration – is used. The percentage of day, evening and night flights and the distribution of aircraft types as stated in Section 4 are also applied to this number of movements. For the second step, the main objectives are to choose suitable routes for the four SIDs and to allocate all the 337 flights featuring two different aircraft types and flown in different time periods (day, night or evening) to these routes. The details of the 337 flights under consideration can be seen in the reference case in Table 1. The optimization problem in step 2 features 28 design variables, *viz.* 4 variables for route selection and 24 variables for aircraft allocation (which are the result of the consideration of four SIDs, two aircraft types and three different operational time periods). By looking at the distribution of the population in Fig. 1, it can be seen that the LUNIX SID is the route that causes less annoyance simply because fewer people live in its direct vicinity. From an airport operational perspective, therefore, this route should accommodate as many flights as possible. However, based on the reference data in Table 1, only around 50% of the flights departing towards the IVLUT intersection are recorded on this route. This is the consequence of the fact that the LUNIX SID intersects the KUDAD SID, limiting the use of the LUNIX SID. Therefore, to include this issue in the allocation problem, we assume that the KUDAD and LUNIX SIDs can handle only 50% of the total number of movements, while RENDI and LEKKO can handle up to 80%. It should also be noted that this assumption can be easily adapted in the framework when the actual operational capacity becomes available. For the integrated optimization problem (the one-step approach), the design variables are the combined variables of both steps in the two-step approach, except for the 4 variables related to route selection. This leads to an optimization problem with 44 variables in total.

---

[2] In this study, the vertical profile is set as follows: from the start to an altitude of 800 ft, full take-off thrust is applied, and $V_2$+10 kts is maintained. After reaching 800 ft, thrust is cut back to climb thrust, and the aircraft is accelerating to $V_{clean}$ whilst continuing a moderate climb. After retracting the flaps, the aircraft maintains climb thrust at a moderate climb gradient until the final conditions are met. At that point, thrust is reduced to maintain these conditions.

To evaluate the reliability and efficiency of both approaches, the results based on the reference case and those derived from the optimal allocation of aircraft movements based on the current SID routes are also estimated and provided. It should be noted that since the studies that address the problem as presented in this paper are not available in the literature, only the results derived from the reference case are used for comparison purposes. All results are shown in Fig. 5. At first glance, it can be seen that both approaches offer solutions that are significantly better than those derived from the reference cases in terms of both the number of people annoyed and fuel burn. A comparison of the result obtained by the reference case and those obtained by the optimal allocation based on the current SID routes indicates that the optimal allocation of flights has a positive influence on the reduction of noise annoyance and fuel burn. When also including the optimized routes, it can be observed that the solutions obtained by both the one-step and two-step approaches are quite close together. Although the one-step approach produces some solutions that are slightly better than those obtained by the two-step approach, the differences between them remain very small. It should be noted that the reason why the one-step approach slightly outperforms the two-step approach is that in the one-step approach, the interaction of the design of optimal routes and the allocation of aircraft on these routes takes place simultaneously, allowing to directly deal with the coupling that exists between these two sub-problems. Meanwhile, in the two-step approach, there is no simultaneous interaction between these sub-problems because these sub-problems are essentially solved sequentially.

Regarding the computational cost, the two-step approach is far more efficient than the integrated approach. Specifically, to obtain these results, the one-step approach requires about 35 hours (h) CPU time, while the calculation time of the two-step approach is only 18 h, mainly spent on the design of optimal routes in Step 1. Clearly, the computational cost is a major restriction of the integrated approach and may be limiting for large applications. Also, the obtained results are less flexible when in practice the number of flights is changed, and a reallocation of flights to the routes is required. Meanwhile, since the two-step approach solves the problem via separate steps, the complexity of the optimization problems has been decreased significantly. In addition, with the optimal routes obtained in the first step, the allocation of flights can be easily reevaluated at a computational cost of around 30 minutes CPU time when a reallocation of flights is requested. Furthermore, the computational cost of the allocation problem can also be further improved by using parallel computing with multiple cores or cluster computing. This is because the evaluation of objective functions in the optimization algorithm is independent.

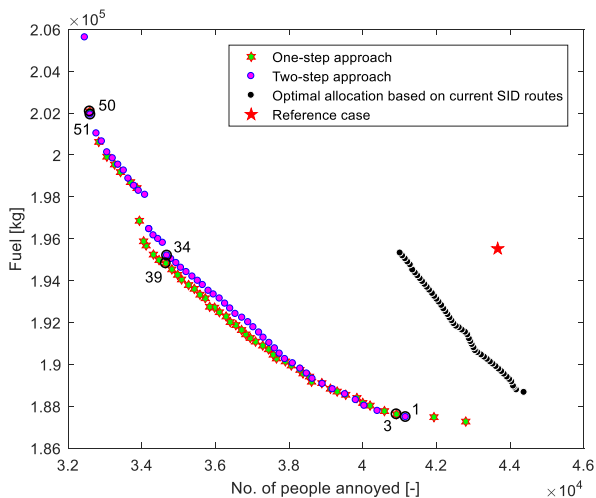

Fig. 5. Comparison of solutions obtained by the one- and two-step approaches and the reference case.

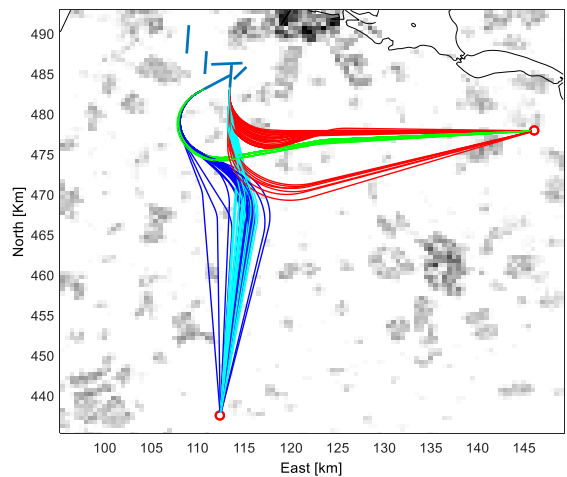

Fig. 6. Optimal ground tracks obtained by the one-step approach (the colors indicate the routes of different SIDs).

Fig. 6 shows the optimal ground tracks obtained by the one-step approach, while those of the two-step approach are displayed in Fig. 7. A comparison of the ground tracks in Fig. 6 and Fig. 7b shows that they are quite similar for both SIDs. From Fig. 6, it can also be observed that the integrated approach creates more route options than the two-step approach. As can be seen in Fig. 5, however, the difference in objective functions between both approaches is moderate.

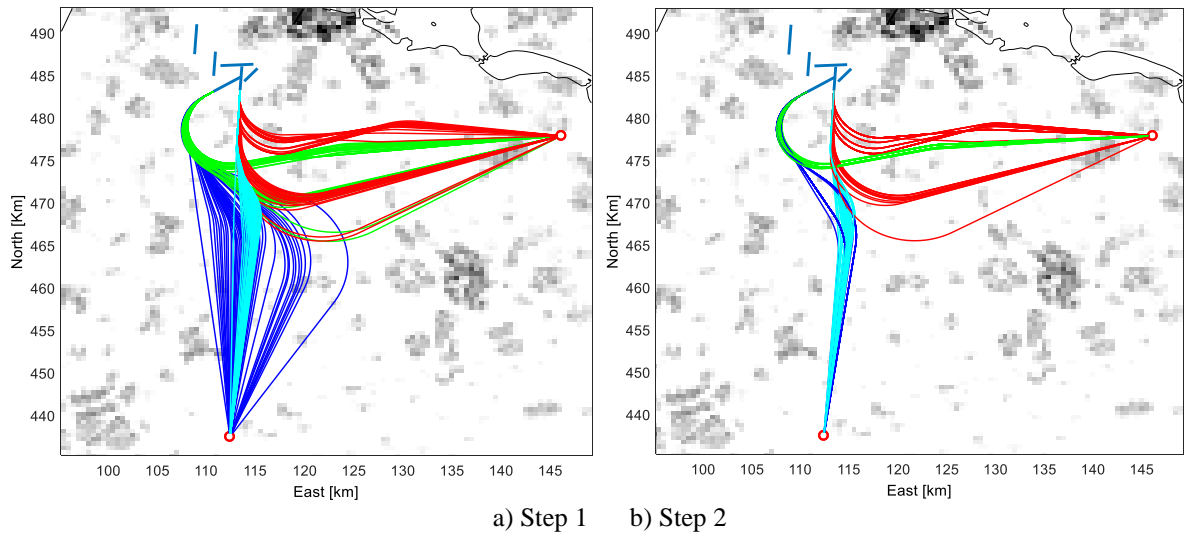

a) Step 1     b) Step 2
Fig. 7. Optimal ground tracks obtained by the two-step approach.

For a more detailed evaluation of the optimal results, the optimal routes of representative solutions (3, 39 and 51 for the one-step approach, and 1, 34 and 50 for the two-step approach) as labelled in Fig. 5 are presented in Fig. 8. The reason for selecting these solutions is that they effectively represent the different aspects of noise and fuel preference, while they closely located on the Pareto fronts. The details of aircraft allocation of these solutions are provided in Table 1. From Fig. 8, it can be seen that all the routes tend to be close together, which reduces the width of the $L_{den}$ contour areas, and consequently, may result in a narrow corridor of high noise exposure between major communities. It is also observed from the figure that the solutions acquired by both approaches exhibit the same trend. It should be noted that the aim of choosing these representative solutions is just to give an overview of the optimal solutions, and does not mean that they are solutions to be recommended for authorities or policymakers. The selection of solutions should be based on their preference, such as noise impact, fuel consumption or the trade-off between them. Also, other criteria associated with each solution, such as sleep disturbance, the fair distribution of noise over population and airspace capacity should be considered.

Therefore, to select suitable solutions, deeper analyses and selection methods should be studied. However, they are not covered in this work and hence are left for further research.

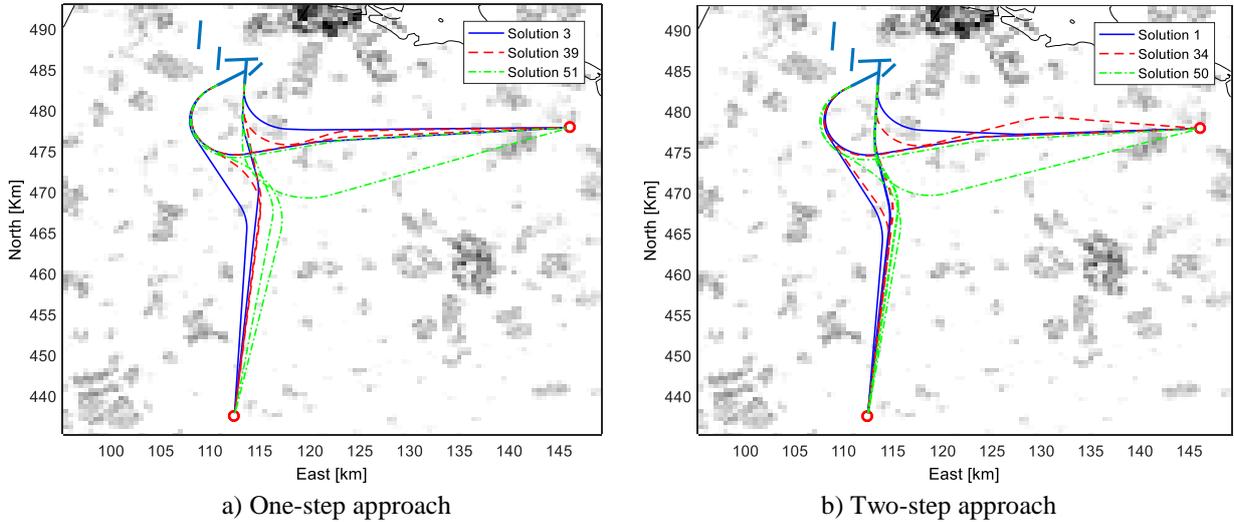

a) One-step approach  b) Two-step approach

Fig. 8. Optimal ground tracks of the representative solutions obtained by the one-step and two-step approaches.

In a comparison of the allocation of aircraft to the presented routes, Table 1 shows the difference in the distribution of aircraft types and the number of flights to the routes between the solutions and the reference case. From Table 1, it can be seen that the general distribution of movements is quite similar for both approaches. Solutions with a lower number of annoyed people (i.e. solutions 50 and 51) tend to prefer the LEKKO SID for the large number of B738s, whereas the B773s are mostly using the KUDAD SID. This is mostly because the limited number of B773s cause less noise, but more importantly, burn more fuel and as such prefer the shorter KUDAD route. Another observation from these results is that although the total number of aircraft using the LUNIX or RENDI SID is quite similar, the LUNIX SID is clearly preferred for evening and night flights, for either aircraft type. This is again due to the balance of fuel and noise. The LUNIX SID causes less annoyance but is a longer route than RENDI. As a result of the weighting factors in Eq. (5) for evening and night flights, moving specifically these flights to LUNIX has a significant positive influence on the noise impact, whereas the overall increase in fuel burn is limited. Table 2 presents the comparison of specific criteria obtained by these solutions and the reference case. It should be noted that while Table 1 displays the changes in the number of flights on each route, the associated changes in the flight schedule and flight delays are still unknown at this stage, due to the fact that the aircraft sequencing problem has not yet been taken into account. These evaluations will be further studied in future work.

Table 1. Optimal aircraft allocation of the representative solutions and the reference case.

| Aircraft allocation | | One-step approach | | | Two-step approach | | | Reference case |
|---|---|---|---|---|---|---|---|---|
| | | Sol. 3 | Sol. 39 | Sol. 51 | Sol. 1 | Sol. 34 | Sol. 50 | |
| B738 | KUDAD | [34,0,0]* | [7,0,0] | [5,0,0] | [38,0,1] | [5,0,0] | [5,0,0] | [36,10,0] |
| | LEKKO | [32,10,21] | [59,10,21] | [61,10,21] | [28,10,20] | [61,10,21] | [61,10,21] | [30,0,21] |
| | LUNIX | [15,3,25] | [50,24,25] | [52,24,25] | [18,0,25] | [47,24,25] | [45,24, 25] | [53,2,25] |
| | RENDI | [108,21,0] | [73,0,0] | [71,0,0] | [105,24,0] | [76,0,0] | [78,0,0] | [70,21,0] |
| B773 | KUDAD | [17,3,4] | [17,3,0] | [17,3,0] | [17,2,1] | [17,2,0] | [17,3,0] | [10,3,0] |
| | LEKKO | [0,0,1] | [0,0,5] | [0,0,5] | [0,1,4] | [0,1,5] | [0,0,5] | [7,0,5] |
| | LUNIX | [0,0,0] | [0,0,6] | [1,0,6] | [0,0,0] | [0,6,6] | [1,6,6] | [13,1,6] |
| | RENDI | [31,6,6] | [31,6,0] | [30,6,0] | [31,6,6] | [31,0,0] | [30,0,0] | [18,6,0] |

[…]*: Number of flights [day, evening, night]

Table 2. Comparison of the criteria of the representative solutions and the reference case.

| Criteria | One-step approach | | | Two-step approach | | | Reference case |
|---|---|---|---|---|---|---|---|
| | Sol. 3 | Sol. 39 | Sol. 51 | Sol. 1 | Sol. 34 | Sol. 50 | |
| No. of people annoyed | 40918 | 34654 | 32588 | 41167 | 34687 | 32606 | 43759 |
| Fuel (ton) | 187.62 | 194.83 | 202.08 | 187.48 | 195.20 | 201.93 | 196.26 |
| Distance (km) | 13372 | 14412 | 15394 | 13359 | 14414 | 15378 | 14327 |
| Flight time (h) | 28.82 | 30.81 | 32.76 | 28.79 | 30.82 | 32.78 | 30.66 |

As can be expected, the integrated approach, in general, provides the best solutions, as can be observed from Fig. 5. However, the differences with the two-step approach are very small indeed, and the latter approach clearly outperforms the former approach in terms of computational effort and flexibility. It can, therefore, be concluded that the two-step approach provides a valid means to combine the optimal routing and allocation problems.

**4.3. 3D optimization case**

In this section, the performance of both approaches is further evaluated for a more complicated optimization problem, where the vertical profile along the route is also considered in the trajectory optimization in Step 1. This case study is also used to assess the potential benefits of optimized vertical profiles in terms of reducing noise impact, which was not considered in previous studies (Braakenburg et al., 2011; Hartjes et al., 2014; Song et al., 2014).

By introducing new design variables for the vertical profile, the number of design variables of the optimization problem in Step 1 is increased significantly. While the ground track of the route is defined in the same way as in the previous section, the vertical part is subdivided into 10 segments and parameterized following the study in Hartjes and Visser (2016), resulting in an addition of 18 design variables. By optimizing the vertical profiles for two different aircraft types concurrently, the number of design variables for the optimization problem in this step now totals 41. Meanwhile, the optimization problem in Step 2 is kept the same as the one in the previous section. The number of design variables of the integrated optimization problem is now 186 in total.

Fig. 9 compares the solutions obtained by the two different approaches and the reference case. As can be seen in Fig. 9, there is a significant reduction in both fuel burn and the number of people annoyed when the vertical profiles are also optimized. This shows that the vertical profile has an important influence on the objective functions when minimizing for fuel burn and noise impact. Comparing the solutions obtained by both approaches reveals that the two-step approach provides solutions that are better than those of the one-step approach. It should be noted that to achieve these results, the two-step approach spent 26 h CPU time in total (of which only 0.58 h is used for the allocation problem with 443 iterations). Meanwhile, the one-step approach took 73 h CPU time after reaching the maximum number of iterations of 1000 that was set for the algorithm. This means that the solutions of the integrated problem still did not yet reach convergence. Although the integrated approach should theoretically always identify better results than the proposed two-step approach, it is clear from these results that the required computational effort is just too high. The results of the integrated solution in 3D are, therefore, not analyzed any further in this section.

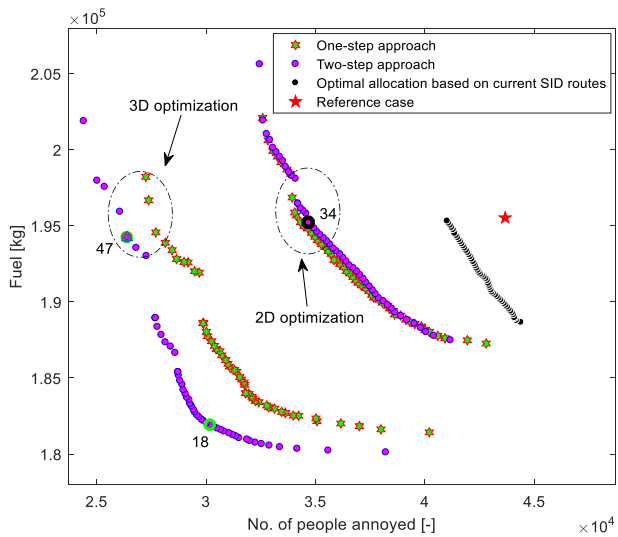
Fig. 9. Comparison of objectives obtained by the one-step and two-step approaches and the reference case

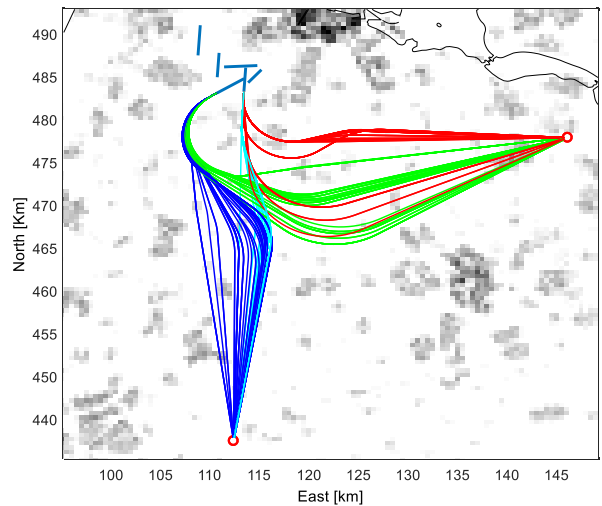
Fig. 10. Optimal ground tracks obtained by the two-step approach.

In an effort to provide a more detailed analysis of these solutions, some representative solutions (as labelled in Fig. 9) for both 2D and 3D approaches and the reference case are selected. The specific performance metrics obtained by these solutions are presented in Table 3. With almost the same performance associated to fuel burn, it can be seen from Table 3 that the 3D solution 47 offers a reduction of up to 23.9% in the number of people annoyed compared to the 2D solution 34, and 39.6% compared to the reference case. Although there is a slight increase of 6.1% in the total distance as a result of avoiding populated regions, owing to the optimized vertical profile the fuel burn has still been reduced. Along the Pareto front of the 3D solutions, we can also identify solutions that outperform both the 2D solutions and the reference case regarding all defined metrics. An example of this is solution 18. Besides achieving a 7.3 % reduction in fuel burn, solution 18 also provides good performance in the number of people annoyed, total distance and flight time, showing a decrease of 31%, 5.7% and 6.7%, respectively, compared to the reference case.

Table 3. Comparison of criteria of the 2D and 3D representative solutions and the reference case.

| Criteria | Reference case | 2D optimization (solution 34) | 3D optimization (solution 47) | (solution 18) |
|---|---|---|---|---|
| No. of people annoyed | 43759 | 34687 | 26400 | 30189 |
| Fuel (ton) | 196.26 | 195.20 | 194.23 | 181.90 |
| Distance (km) | 14327 | 14414 | 15207 | 13511 |
| Flight time (h) | 30.66 | 30.82 | 32.16 | 28.60 |

Fig. 10 shows the ground tracks obtained by the two-step approach, while the ground tracks and the vertical profiles of the solutions 18, 34 and the reference case are given in Fig. 11 and Fig. 12, respectively. From Fig. 12, it can be seen that there is a significant difference between the optimized vertical profiles and the reference profile in the first part of the routes. In the optimized profiles, the emphasis lies more on acceleration in the initial part of the trajectory, whilst the standard profiles try to keep a balance between speed and altitude. The increased acceleration featured in the optimized profile allows for an earlier flap retraction and, in general, better performance in terms of fuel burn. In addition, the low altitude flight at a higher airspeed reduces the noise impact. The reason for this is twofold. Firstly, the higher airspeed leads to a lower exposure time and hence lower $SEL$ and $L_{den}$ values. In addition, the low altitude flight – although the noise exposure directly below the flight path is higher – leads to a lower noise exposure astride the trajectory, as the lateral attenuation losses are significantly higher. As a result, the number of annoyed people has significantly reduced. The illustration of this can be seen in Fig. 13, where the difference of $L_{den}$ noise contour areas caused by solutions 18, 34 and the reference case, and the number of people annoyed on each grid cell are clearly illustrated.

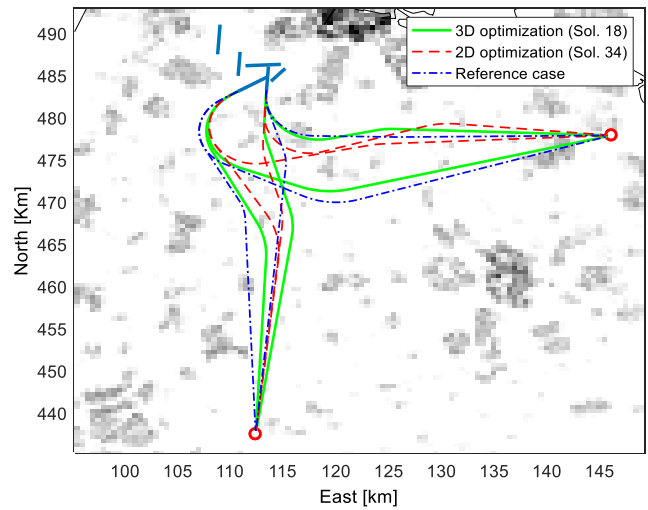

Fig. 11. Comparison of solutions obtained by the 2D and 3D optimization and the reference case.

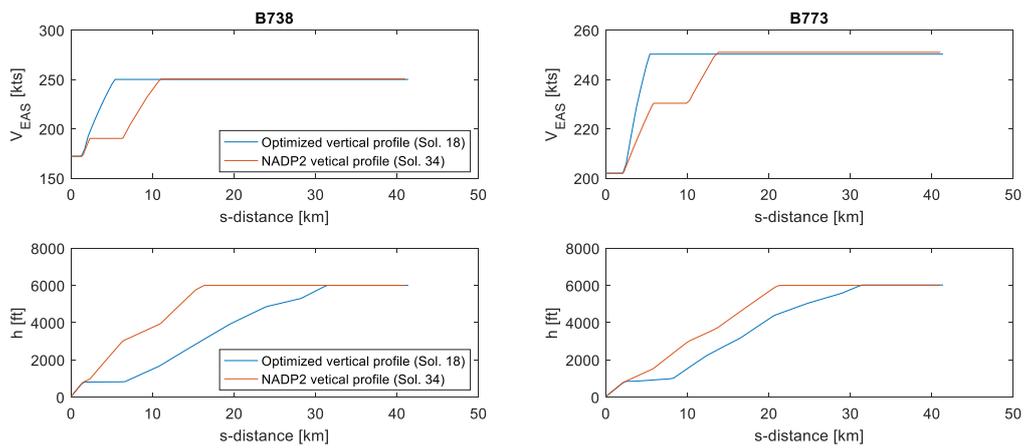

a) Route KUDAD

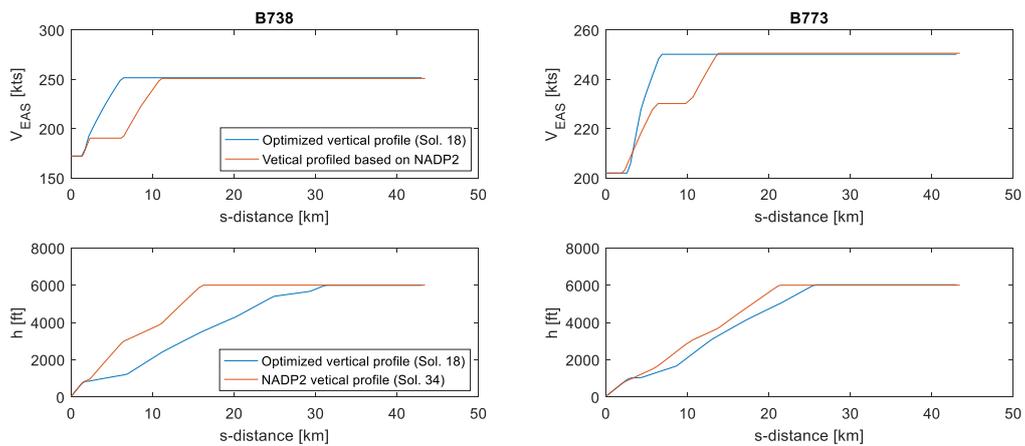

b) Route LEKKO

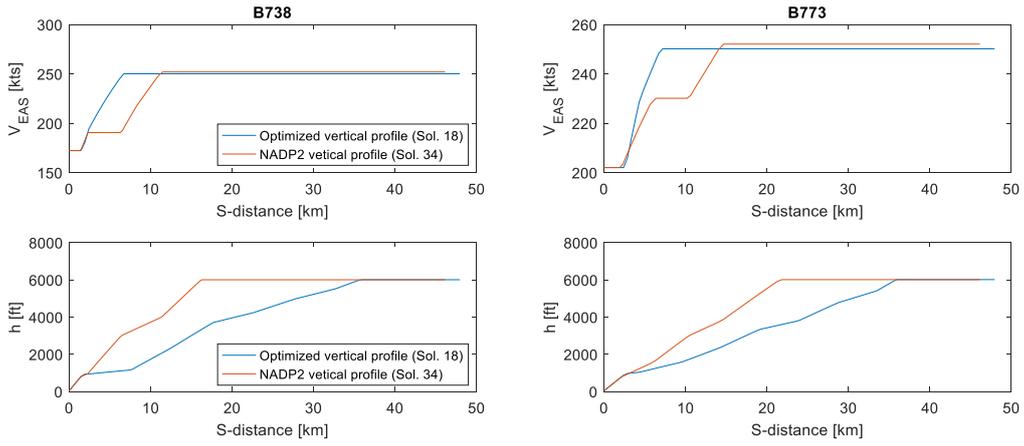

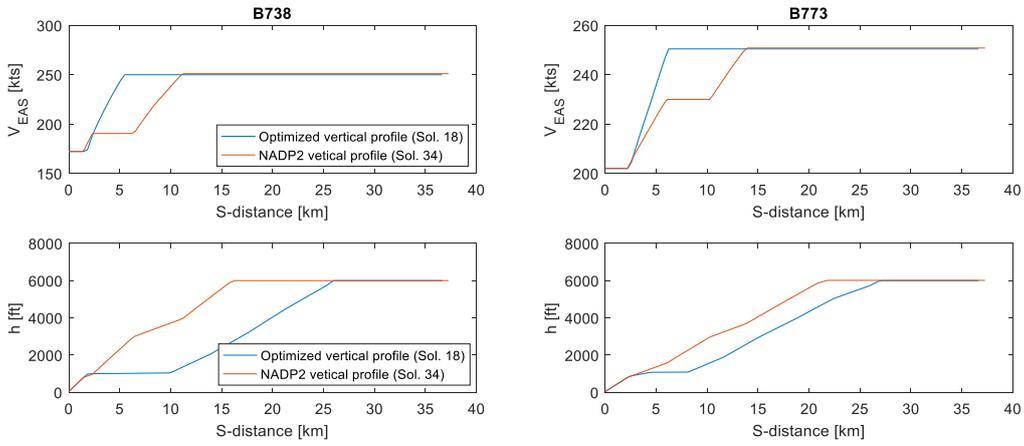

c) Route LUNIX

d) Route RENDI

Fig. 12. Vertical profiles of the 3D solution (solution 18) and those based on the reference profile (solution 34).

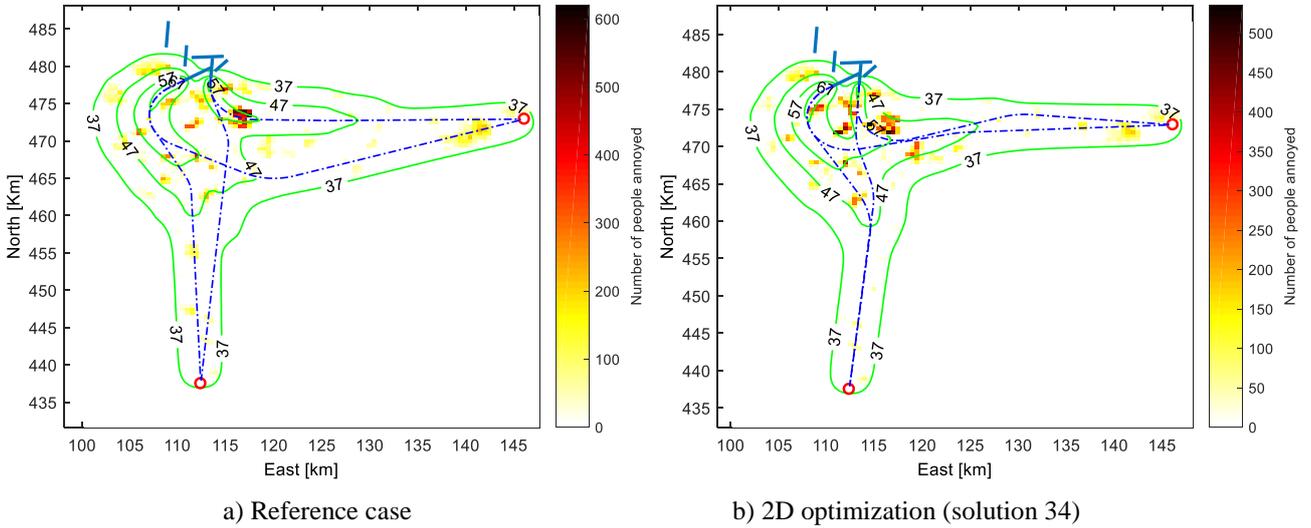

a) Reference case

b) 2D optimization (solution 34)

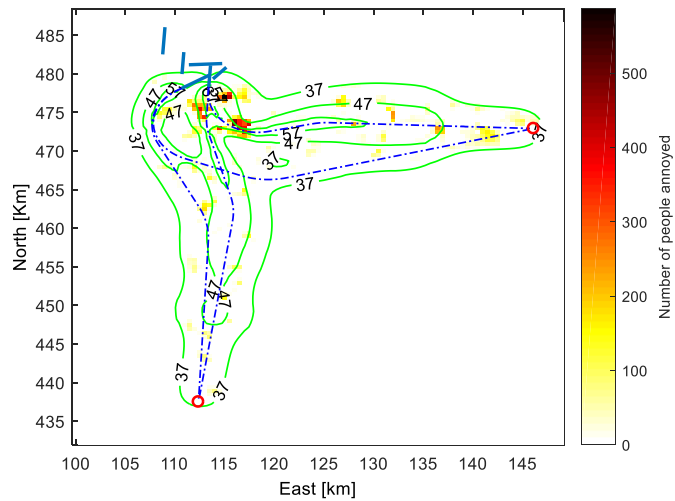

c) 3D optimization (solution 18)

Fig. 13. Illustration of $L_{den}$ and the number of people annoyed caused by the representative solutions and the reference case.

## 5. Conclusion

In this paper, we have presented and validated a two-step optimization framework for the design of optimal aircraft departure routes and the distribution of aircraft movements over these routes. Firstly, to explore the potential of reducing noise and fuel burn for each standard instrument departure (SID), the multi-objective trajectory optimization problem is formulated and solved in the first step. Secondly, the obtained sets of optimal routes are then used as the inputs for the optimization problem in the second step, where the selection of optimal routes for SIDs and the assignment of flights among these routes are optimized simultaneously. The reliability of the framework has been validated through the analysis of the employed noise criterion and the solution of the integrated optimization problem in a single step. A case study at Amsterdam Airport Schiphol (AMS) in The Netherlands has been used to assess the efficiency and capability of the proposed framework.

The numerical results indicate that the present framework is reliable and able to provide solutions which can significantly reduce noise annoyance and fuel consumption. Moreover, the obtained results have shown that from a theoretical perspective, the one-step approach can, in principle, fully exploit the potential of noise and fuel reduction by solving the integrated optimization problem. However, the computational cost and the complexity associated to the integrated optimization problem are prohibitively large. Furthermore, this approach is also less flexible with respect to changes in the number of flights when a reallocation of flights is demanded or new routes or runways are considered. Although the integrated problem can theoretically lead to better results, the two-step framework proposed in this study has proven to be a valid and viable alternative, able to overcome the mentioned issues.

In view of the attained favorable results, the framework appears to be suitable for extension to other applications such as the design and allocation of aircraft arrival routes, and the combined problem of departure and arrival routes. Moreover, an application with larger scale and scope, for instance, an entire airport, will also be considered in follow-on studies. In the current study, the actual influence of optimal allocation solutions on the airspace capacity and aircraft sequencing problem has not yet been considered. This issue will also be addressed in further research.

**Acknowledgments**

The authors would like to thank the editor and anonymous reviewers for their constructive, helpful and valuable comments and suggestions.

**Appendix A. Notation**

| | |
|---|---|
| $\dot{V}_{TAS}$ | The derivative of the true airspeed with respect to time, $V_{TAS}$ |
| $\dot{V}_{EAS}$ | The derivative of the equivalent airspeed with respect to time, $V_{EAS}$ |
| $\dot{s}$ | The derivative of the ground distance flown with respect to time, $s$ |
| $\dot{h}$ | The derivative of the altitude with respect to time, $h$ |
| $\dot{W}$ | The derivative of the aircraft weight with respect to time, $W$ |
| $\dfrac{\partial \rho}{\partial h}$ | The derivative of the ambient air density $\rho$ with respect to altitude $h$ |
| $\dot{m}_0$ | Fuel flow |
| $T$ | Thrust |
| $D$ | Drag |
| $\gamma$ | Flight path angle |
| $g_0$ | Gravitational acceleration |
| $\rho$ | Air density at sea level |
| $\rho_0$ | Ambient air density |
| $\%PA$ | The percentage of people annoyed |
| $L_{den}$ | Day-evening-night cumulative noise metric |
| $N_r$ | Total number of departure routes (SIDs) |
| $N_{at}$ | Total number of aircraft types |
| $SEL_{ki}$ | Sound exposure level resulted from aircraft type $i$ on route $k$ |
| $w_{den}$ | Weighting factor |
| $a_{ki}$ | Number of aircraft type $i$ operating on route $k$ |
| $T$ | Considered time period |
| $N_{pa}(\mathbf{d})$ | Total number of people annoyed in the first step |
| $T_{fuel}(\mathbf{d})$ | Total fuel burn in the first step |
| $\mathbf{d}$ | Vector of design variables |
| $\mu_i(t)$ | Bank angle of aircraft type $i$ |
| $R$ | Turn radius |
| $\mu_{max}$ | Maximum permissible value of $\mu$ |
| $a_i$ | Number of aircraft type $i$ |
| $fuel_i(\mathbf{d})$ | Fuel burn of aircraft type $i$ |
| $\mathbf{r}$ | Vector of design variables of departure routes |
| $\mathbf{O}_k$ | Set of optimal routes for the SID $k$ |
| $\mathbf{a}$ | Design variable vector of the aircraft allocation |
| $a_{itk}$ | Number of aircraft type $i$ at time $t$ on route $k$ |
| $\mathbf{SD}_s$ | Vector containing the number of departure routes $l$ having the same directional point $s$ |
| $T_{at,its}$ | Total number of aircraft type $i$ at time $t$ sent to departure routes having the same terminal point $s$ |
| $\bar{N}_{f,k}$ | Upper bound of the number of movements that route $k$ can handle in a certain period of time |
| $\bar{a}_{itk}$ | Upper bound of the number of aircraft type $i$ on route $k$ at time $t$ |
| $N_{pa}(\mathbf{r},\mathbf{a})$ | Total number of people annoyed in the second step |
| $T_{fuel}(\mathbf{r},\mathbf{a})$ | Total fuel burn in the second step |
| $fuel_{ik}(r_k)$ | Fuel burn of aircraft type $i$ on route $r_k$ |